\RequirePackage{fix-cm}
\documentclass[smallextended]{svjour3}        
\smartqed  
\usepackage{graphicx}
\begin{document}

\title{Stability of the intrinsic energy vanishing in the Schwarzschild metric under a slow rotation}


\titlerunning{Stability of the intrinsic energy vanishing}        
\author{Juan M. Aguirregabiria \and Ramon Lapiedra \and Juan Antonio Morales-Lladosa}


%
\institute{Juan M. Aguirregabiria \and Ramon Lapiedra* \and Juan Antonio Morales--Lladosa* 
\at Theorethical Physics,  University of the Basque Country (UPV/EHU), P. O. Box 644, 48080 Bilbao, Spain \at *Departament
d'Astronomia i Astrof\'{\i}sica,
\\Universitat de Val\`encia, E-46100 Burjassot, Val\`encia, Spain.\\
Tel.: +34-96-3543066\\
Fax: +34-96-3543084\\
\email{juanmari.aguirregabiria@ehu.es}\\
\email{ramon.lapiedra@uv.es}\\
\email{antonio.morales@uv.es}}
\date{Received: date / Accepted: date}

\maketitle

\begin{abstract}
The linearized Kerr metric is considered and put in some Gauss coordinates which are further {\em intrinsic} ones. The linear and angular 4-momenta of this metric are calculated in these coordinates and the resulting  value is just zero. Thus, the global vanishing previously found for the Schwarzschild metric remains linearly stable under 
slow rotational perturbations of this metric.

\end{abstract}

\keywords{Space-time momenta and asymptotic
flatness \and Schwarzschild and Lense-Thirring 
metrics \and  Weinberg complex}
\PACS{04.20.Cv \and 04.20.-q}


\section{Introduction}
\label{intro}

In Relativity, global quantities  such as energy are usually defined from initial data on a space-like 3-dimensional slice. In the  asymptotically Minkowskian case, these quantities make sense  provided that a suitable  falloff rate of the metric and extrinsic curvature of the above slice are fulfilled (see, for instance, \cite{EricG,Alcubierre} for a general account and more references on the subject).  Of course, different choices of slices might lead to different values of these quantities. Furthermore, for a given slice, the obtained values also depend of the chosen coordinates on the slice because this choice can change the asymptotic falloff rates of the metric components, providing different physical interpretations for different calculated values. 

A more specific approach, largely represented in the literature on the subject (see, for example \cite{AguirreAV-96}) is the one starting from an energy-momentum complex, which is the approach to which we adhere in the present paper. In fact,  for a linearized Kerr metric,  we are planing to construct suitable coordinates systems ($T,\rho,\theta,\Phi$) where all the Weinberg global quantities vanish for a slice $T=T_0$. For a different and interesting approach to this issue in General Relativity see  \cite{Cooperstock-2000,Cooperstock-Dupre}.

For instance, in the Schwarzschild space-time the energy associated with the static observers is the Schwarzschild mass parameter  \cite{EricG}.  Nevertheless,  
for a Lema{\^{\i}}tre congruence of observers \cite{Elmestre} the energy vanishes (cf \cite{EricG,Zannias,Chrusciel}). In fact, the former observers are non-geodesic 
meanwhile the later are non-rotating and radially free-falling from the rest at the spatial infinity (in the specific sense that these observers are asymptotically at rest at this spatial infinity) 
and their constant coordinate time slices are flat. In a precedent paper \cite{Schwarzschild-Creatable} 
we have justified  that  these different results for the energy are both physically acceptable and compatible  with the respective underlying observer kinematics. 

Furthermore, in  \cite{Schwarzschild-Creatable}, given the Schwarzschild metric of an ideal non rotating star, two of the present authors have proved that there is a coordinate system in which the linear, $P^\alpha$, and the angular, $J^{\alpha \beta}$,  4-momenta ($\alpha, \beta, ... = 0, 1, 2, 3$) of this metric vanish, its 3-space components $J^{ij}$ ($i, j, ... = 1, 2, 3$) vanishing irrespective of the $J^{ij}$ origin. We have called ``self-creatable'' (``creatable'' for short) the metrics whose such two 4-momenta vanish \cite{Ferrando,Lapiedra-Saez,NewCreatable} in a given coordinate system, since such metrics could perhaps arise from a vacuum quantum fluctuation \cite{Albrow,Tryon,Vilenkin-1985}. Needless to say that this ``creation'' is a very speculative subject, though, in our opinion, one worth to be followed on searching in particular for possible inner or observational inconsistencies (see \cite{LTB-14}).

Of course, strictly speaking, the adjective ``creatable'' applied to a given metric in General Relativity could be reduced to a mere short for saying that this metric has vanishing {\em intrinsic} linear and angular 4-momenta.

Thus, given a metric, its two associated 4-momenta, $P^\alpha$ and $J^{\alpha \beta}$, are not unique depending on the type of coordinates used. Furthermore  they are dramatically dependent of this type of coordinates. Thus, we must specify that these vanishing $P^\alpha$ and $J^{\alpha \beta}$ values of the Schwarzschild metric, that we considered in  \cite{Schwarzschild-Creatable}, were the ones called {\em intrinsic} 4-momenta values, that is, the ones calculated in {\em intrinsic} coordinates, which, in the particular case of Schwarzschild metric, become coordinates adapted  to the spherical symmetry of the metric and to a family of radially free-falling, synchronized, observers at rest at the 3-space infinity. The synchronous observers are also considered as the natural ones in \cite{Cooperstock-Dupre} in order to calculate a sound energy.

This vanishing of the {\em intrinsic} $P^\alpha$ and $J^{\alpha \beta}$ was obtained directly when the source radius, $r_1$ (the radius of the above ideal star), was larger than the Schwarzschild radius $r_0 = 2m$, where $m$ is the Schwarzschild mass (we take the gravitational constant, $G$, and the speed of light, $c$, equal to 1). In the opposite case, $r_1 < r_0$, that is, in the case of a gravitational collapse, we only attain this vanishing result by using what we call the more than quasi-local algorithm. The use of this algorithm resorts to calculate the {intrinsic} values of
$P^\alpha$ and $J^{\alpha \beta}$ for a given constant time, say $T_0$, in an {\em intrinsic} coordinate system associated to $T_0$, such that the associated {\em intrinsic} system changes when we change $T_0$.

The fact that we obtained the vanishing of the {\em intrinsic} values of $P^\alpha$ and $J^{\alpha \beta}$, both when $r_1 > r_0$, and alternatively when the star collapses to a black hole, $r_1 < r_0$, is in accordance with a minimal consistency, since, starting from $r_1 > r_0$ we could imagine an ideal spherical collapse of our ideal star. We mean a radial collapse where neither any external matter-energy is infalling, nor any inner matter-energy,  including any gravitational radiation, is ejected. Then, we could conjecture that these vanishing values of $P^\alpha$ and $J^{\alpha \beta}$ should remain constant all along the entire collapsing process, and since they were zero initially, they should remain zero for the resulting black hole.

On the other hand, in the cited article  \cite{Schwarzschild-Creatable}, we asked ourselves whether the obtained ``creatable'' character of the Schwarzschild metric could be unphysical, in the sense that the result were unstable. That is, we asked ourselves whether this vanishing of the {\em intrinsic} values of $P^\alpha$ and $J^{\alpha \beta}$ could disappear if we considered a slowly rotating star, instead of the non rotating considered up to then.

The aim of the present article is to prove, in the case of such a rotating star, that there are {\em intrinsic} values of $P^\alpha$ and $J^{\alpha \beta}$ all them vanishing. Therefore, the rotating metric considered, i. e., the Lense-Thirring metric \cite{Mashhoon}, is {\em creatable}, which means that the previously found creatability of the Schwarzschild metric in \cite{Schwarzschild-Creatable} is a stable result against any linear Kerr generalization of the Schwarzschild metric. We remark that our result only concerns the stability under a one-parameter family of small perturbations modeling a slow star rotation.

The plan of the rest of the paper is as follows: In Section \ref{sec-2} we write, in intrinsic coordinates, the linearized Kerr metric in its angular momentum parameter, after defining what {\em intrinsic} coordinates are.  Finally, 
in Section \ref{sec-3}, we calculate the linear and angular 4-momenta of this linearized metric, these 4-momenta being the ones derived from the Weinberg pseudo-tensor. We obtain vanishing 4-momenta, the angular 3-momentum vanishing irrespective of the momentum origin,  and draw the final conclusion of this 4-momenta vanishing.

%
%
%


\section{Linearized Kerr metric in {\em intrinsic} coordinates}
\label{sec-2}

An extensive analysis  on the linear and angular momenta of  the Kerr metric has been carried out in \cite{AguirreAV-96}, by using several suitable coordinate systems and different pseudotensorial prescriptions. Here, our main subject concerns the stable character of the creatability of the Schwarzschild geometry, and then, we limit our considerations to the Kerr linearized case  in Gauss coordinates. 

Using Boyer-Lindsquist coordinates, the Kerr metric to first order in the parameter $a$, that is in the angular momentum of the source per unit of mass, reads
\begin{equation}\label{KerrLT-metrica}
 ds^2= - (1 - \frac{r_0}{r})dt^2 +  \frac{dr^2}{1 - \frac{r_0}{r}} + r^2 d \sigma^2 - 2 a \frac{r_0}{r}  \sin^2 \theta \, dt \, d\phi,
\end{equation}
with $d \sigma^2 \equiv d \theta^2 + \sin^2 {\hspace{-0.7mm}} \theta \, d \phi^2$, and where $a/r \ll 1$ and $r_0$ is the Schwarzschild radius.

As it is well known (see for instance \cite{Weinberg}), modulo ${\cal O}(r_0^2/r^2)$, when transformed into harmonic coordinates, this linearized Kerr metric becomes the Lense-Thirring metric \cite{Mashhoon} which describes the metric of a slowly rotating sphere far away enough, that is, for small values of  $a$ and $r_0/r$, although for convenience, all along the paper, we have (unnecessarily) maintained the exact expression in $r_0/r$ for the Schwarzschild part of the metric line element, even when we have in mind the corresponding Lense-Thirring metric. 

In order to calculate the {\em intrinsic} values of $P^\alpha$ and $J^{\alpha \beta}$, for this metric (\ref{KerrLT-metrica}), we need previously transform it into {\em intrinsic} coordinates  \cite{Schwarzschild-Creatable}, which in the case of an asymptotically Minkowskian metric, are by definition (signature $+2$):

(a) Gauss coordinates, that is $g_{00} = -1$, $g_{0i} = 0$, at least, fast enough, at the 3-space infinity, $r \to \infty$.

(b) Fast enough, asymptotically rectilinear coordinates at the 3-space infinity.

(c) Such that the linear 3-momentum $P^i$, and the angular 3-momentum $J^{ij}$, vanish,  the last one irrespective of its origin. 

As it is well known \cite{Weinberg}, Gauss coordinates are associated to free falling observers, and on the other hand it has been proved \cite{NewCreatable} that (c) can always be fulfilled 
jointly with (a) and (b).

Then, let us change to the new coordinates $(T, R, \theta, \Phi)$, given by the transformation
\begin{equation} \label{canvi}
t = T - \, r_0 \, f(r), 
\qquad r^{3/2} = \frac{3}{2} \sqrt{r_0} (R - T), \qquad \phi = \Phi - \frac{a}{r_0} \, \tilde{f}(r) - a \,  r_0 \,  h(R), 
\end{equation}
with
\begin{equation} \label{fr}
f(r) \equiv 2 \sqrt{\frac{r}{r_0}} + \ln \Big|\frac{\sqrt{r} - \sqrt{r_0}}{\sqrt{r} + \sqrt{r_0}}\Big|.
\end{equation}
\begin{equation} \label{fr-tilde}
\tilde{f}(r) \equiv 2 \sqrt{\frac{r_0}{r}} + \ln \Big|\frac{\sqrt{r} - \sqrt{r_0}}{\sqrt{r} + \sqrt{r_0}}\Big|.
\end{equation}
and where the function $h(R)$ must go fast enough to zero as $R \to \infty$,  in order to ensure that the metric fulfills the above condition (b) 
in a suitable coordinate system, where we will fit $h$ such that $P^\alpha$ and $J^{\alpha \beta}$ vanish.

Neglecting terms of order ${\cal O}(a^2)$, the metric (\ref{KerrLT-metrica}) becomes 
\begin{equation}\label{KLT-Lemaitre}
 ds^2 = - dT^2 +  \frac{r_0}{r} dR^2+ r^2 (d \theta^2 + \sin^2 \theta \, d\Phi^2) - 2 a \frac{r_0}{r} (1 + r^3 h')  \sin^2 \theta \, dR \, d\Phi,
\end{equation}
with $h' \equiv dh/dR$, showing that the new coordinates are Gauss coordinates, so that the above prescription (a) is fulfilled. However, it can easily be seen that prescription (b) is not fulfilled. In order to fulfill (b), let us introduce a new radial coordinate, $\rho$, defined such that
\begin{equation} \label{rrhoT}
\rho^{3/2} + C = \frac{3}{2} \sqrt{r_0} \, R,  
\end{equation}
where $C$ is an arbitrary constant. A similar change of the radial coordinate was considered in \cite{Oppenheimer-Snyder-1939,Synge-1950,Robertson-Noonan} and used in \cite{Schwarzschild-Creatable}. Then, we will have from (\ref{KLT-Lemaitre})

\begin{equation}\label{KLT-Lemaitre2}
 ds^2 = - dT^2 +  \frac{\rho}{r} d\rho^2+ r^2 (d \theta^2 + \sin^2 \theta \, d\Phi^2) - 2 a  \frac{\sqrt{r_0 \rho}}{r} (1 + r^3 h') \sin^2 \theta \, d\rho \, d\Phi,
\end{equation}
where from (\ref{rrhoT}) and  (\ref{canvi}), we have
\begin{equation}\label{rrhoT2}
r = \big(\rho^{3/2} - \frac{3}{2} \sqrt{r_0} \, T + C \big)^{2/3}.
\end{equation}
Then, we see that for any given $T$ the metric (\ref{KLT-Lemaitre2}) is manifestly the metric of the Minkowski space-time, $M_4$, at the infinity $\rho \to \infty$ 
(remember that function $h$ is assumed to go fast enough to zero when $R \to \infty$), that is,  the Gauss coordinates 
$(T, \rho, \theta, \Phi)$ give asymptotic rectilinear coordinates, $\rho_i$, 
\begin{equation}\label{rho-cartesianes}
\rho_1 = \rho \sin \theta \cos \Phi, \qquad \rho_2 = \rho \sin \theta \sin \Phi, \qquad \rho_3 = \rho \cos \theta.
\end{equation}

In these coordinates $(T, \rho_i)$ the metric (\ref{KLT-Lemaitre2}) reads out ($\delta_{ij}$ is the Kronecker symbol and $n_i = \rho_i / \rho$):
\begin{equation}\label{KLT-Lemaitre2b}
 ds^2 = - dT^2 +  [(\frac{\rho}{r} -\frac{r^2}{\rho^2}) n_i n_j + \frac{r^2}{\rho^2} \delta_{ij}] d \rho_i d\rho_j + 2 a \frac{\sqrt{r_0}}{r \sqrt{\rho}} (1 + r^3 h') n_i d \rho_i (n_2 d\rho_1 - n_1 d \rho_2).
\end{equation}
Having in mind further calculations, it will be convenient to split  (\ref{KLT-Lemaitre2b}) in the form $ds^2 = ds^2_S  + \delta ds^2$, where $\delta ds^2$ is a perturbation of the Schwarzschild metric $ds^2_S$, 
\begin{equation}\label{metric-split}
 ds^2_S \equiv - dT^2 +  [(\frac{\rho}{r} -\frac{r^2}{\rho^2}) n_i n_j + \frac{r^2}{\rho^2} \delta_{ij}] d \rho_i d\rho_j, 
\end{equation}
\begin{equation}\label{metric-split-2}
\delta ds^2 \equiv 2 a \frac{\sqrt{r_0}}{r \sqrt{\rho}} (1 + r^3 h') n_i d \rho_i (n_2 d\rho_1 - n_1 d \rho_2), 
\end{equation}
$\delta ds^2$ fulfilling the asymptotic (b) condition provided that $h'$ goes to zero sufficiently quickly as $\rho \to \infty$.

In the next section, we calculate $P^\alpha$ and $J^{\alpha \beta}$ for this linearized Kerr metric in the coordinates $(T, \rho_i)$, that is, for (\ref{KLT-Lemaitre2b}).


\section{Intrinsic 4-momenta for the linearized Kerr metric}
\label{sec-3}
We take the Weinberg energy-momentum complex \cite{Weinberg} to derive the corresponding general expressions for $P^\alpha$ and $J^{\alpha \beta}$.  
The reasons for this choice have to do with the fact that what we particularly want are the intrinsic values of $P^\alpha$ and $J^{\alpha \beta}$ and, on the other hand, with the fact that this complex comes {\em directly} from the corner-stone of General Relativity: from the Einstein field equations. Here ``directly'' means by a mere reorganization of their different terms (see \cite{Schwarzschild-Creatable}, for details). 
The corresponding expressions for $P^\alpha$ and $J^{\alpha \beta}$ are 3-volume integrals that can be transformed into 2-surface integrals on the boundary of the integration 3-volume, 
provided that we can apply the Gauss theorem to the 3-volume integrals. We can ensure this by giving to the slowly rotating spherical star a finite radius at 
$T = T_0$ such that we could avoid the intrinsic singularity for $r=0$ in the metric (\ref{KerrLT-metrica}). 

Written like such 2-surface integrals, the expressions for $P^\alpha$ and $J^{\alpha \beta}$ become the limits for $\rho \to \infty$ of the following integrals
\begin{eqnarray}
P^0 & = & \frac{1}{16\pi} \int(\partial_j g_{ij} - \partial_i g)
d \Sigma_{i},  \label{energy}\\[3mm]
P^i & = & \frac{1}{16\pi} \int(\dot{g} \delta_{ij} -
\dot{g}_{ij}) d \Sigma_{j},
\label{linear momentum}\\[3mm]
J^{jk} & = & \frac{1}{16\pi} \int(\rho_k \dot {g}_{ij} - \rho_j
\dot {g}_{ki}) d \Sigma_{i},\label{angular momentum}
\\[3mm]
J^{0i} & = & P^i T - \frac{1}{16\pi} \int[(\partial_k g_{kj} -
\partial_j g)\rho_i + g \delta_{ij} - g_{ij}] d \Sigma_{j}, \quad
\label{angular time momentum}
\end{eqnarray}
where the following notation is used: $g \equiv \delta^{ij}g_{ij}$, the dot stands for the partial derivative with respect to $T$, 
and $d \Sigma_{i}$ is the 2-surface integration element. In the present case $d \Sigma_i = n_i \rho^2 d \Omega = n_i \rho^2 \sin \theta d \theta d \Phi$, $n_i \equiv \rho_i/\rho$, and $\partial_i$ denotes the partial derivative with respect to $\rho_i$.


\subsection{Fixing the coordinate system at the slice $T = T_0$}

In order to do this calculation, let us chose  the particular time $T = T_0$ in which we are going to do the calculation. We will have for (\ref{rrhoT2})
\begin{equation}\label{rrhoT3}
r = \big(\rho^{3/2} - \frac{3}{2} \sqrt{r_0} \, T_0 + C \big)^{2/3}.
\end{equation}
Let us take $C = 3 \sqrt{r_0} \, T_0 / 2$ that implies $\rho = r$ on the slice $T = T_0$. Then, the metric (\ref{KLT-Lemaitre2b}) is written as

\begin{equation}\label{K-simple}
 ds^2|_{T = T_0}= ds^2_S |_{T = T_0} + \delta ds^2|_{T = T_0}, 
 \end{equation}
where all the metric components take values  for $T=T_0$, that is
\begin{equation}\label{K-simple2}
 ds^2_S|_{T = T_0} \equiv  - dT^2 + \delta_{ij}  d\rho_i d\rho_j
 \end{equation}
is the part corresponding to the Schwarzschild metric, and
 \begin{equation}\label{delta-ds2}
\delta ds^2|_{T = T_0} \equiv  2 a \sqrt{r_0} \, [1 + \rho^3 \psi(\rho)] \,  \frac{n_i d \rho_i}{\rho^{3/2}}  (n_2 \, d \rho_1 - n_1 \, d \rho_2)
\end{equation}
corresponds to the linearized Kerr correction, where we have put $h'(R) = \psi (\rho)$.

We will still need to calculate $\frac{\partial}{\partial T} (ds^2)$ for $T = T_0$, whose expression, having in mind  (\ref{rrhoT2}) and  (\ref{KLT-Lemaitre2b}),  becomes
\begin{equation}\label{partial-Tds2}
\frac{\partial (ds^2)}{\partial T}\Big|_{T = T_0}  = \frac{\sqrt {r_0}}{\rho^{3/2}} (3 n_i n_j - 2 \delta_{ij}) d \rho_i d \rho_j +  2 a r_0 \, (1-2 \rho^3 \psi) \, 
\frac{n_i d \rho_i}{\rho^{3}}  (n_2 \, d \rho_1 - n_1 \, d \rho_2)
\end{equation}
and then we denote
\begin{equation}\label{partial-Tds2-S}
\frac{\partial (ds^2_S)} {\partial T}\Big|_{T = T_0}  = \frac{\sqrt {r_0}}{\rho^{3/2}} (3 n_i n_j - 2 \delta_{ij}) d \rho_i d \rho_j, 
\end{equation}
\begin{equation}\label{delta-ds2-punt}
\frac{\partial ( \delta ds^2)}{\partial T}\Big|_{T = T_0} =    2 a r_0 \, (1-2 \rho^3 \psi) \, \frac{n_i d \rho_i}{\rho^{3}}  (n_2 \, d \rho_1 - n_1 \, d \rho_2).
\end{equation}

Notice that, in accordance with the definition of intrinsic coordinates given in Section \ref{sec-2},  the choice  $C = 3 \sqrt{r_0} \, T_0 / 2$ fixes a suitable family of intrinsic coordinates $(T, \rho_i)$ in which we are going to prove that both 4-momenta of the metric (\ref{KLT-Lemaitre2b}) vanish for a suitable choice of $\psi$.


\subsection{Obtaining $P^\alpha = 0$ and $J^{ij} = 0$}

As it has been stated in the Introduction, the intrinsic values of $P^\alpha$ and $J^{\alpha \beta}$ for the Schwarzschild metric vanish, its 3-space components, $J^{ij}$, vanishing irrespective 
of the angular 3-momentum origin \cite{Schwarzschild-Creatable}. Let us specify that, in this case,  the corresponding intrinsic coordinates where this vanishing takes place are just the coordinates $(T, \rho_i)$ for the Schwarzschild space-time.

Therefore, in (\ref{energy})-(\ref{angular time momentum}) we can substitute the metric  $g_{ij}$ and its time derivative, $\frac{\partial g_{ij}}{\partial T}$, at $T_0$ (given implicitly by (\ref{K-simple})-(\ref{delta-ds2}), and by (\ref{partial-Tds2}), respectively) by $\delta g_{ij}|_{T = T_0}$ and by $\partial_T (\delta g_{ij})|_{T = T_0}$ (given implicitly by (\ref{delta-ds2}) and by (\ref{delta-ds2-punt})).

Then, let us take the arbitrary function $\psi (\rho)$ such that $2 \psi = \rho^{-3}[1 + {\cal O}(\rho^{-n})]$, $n>0$. This choice leads straightforwardly to
\begin{equation}\label{pTdelta-ds2-zero}
\frac{\partial (\delta ds^2)} {\partial T}\Big|_{T = T_0}  = 0
\end{equation}
modulo terms that fall faster than $\rho^{-3}$ for $\rho \to \infty$.  We will see next that this kind of decay guarantees the vanishing of  $P^{\alpha}$ and $J^{\alpha \beta}$. First of all, from 
(\ref{linear momentum}) and (\ref{angular momentum}), we find easily

\begin{equation}\label{3momenta-zero}
P^i = 0, \quad  J^{i j}= 0.
\end{equation}

Further, according to (\ref{delta-ds2}), $\partial_i (\delta g_{jk})|_{T = T_0} \sim \rho^{-5/2}$, which means that the corresponding limit $\rho \to \infty$ 
of the integral (\ref{energy}) trivially vanishes too, i.e., 
\begin{equation}\label{E0}
P^0 = 0.
\end{equation}
%


\subsection{Obtaining $J^{i0} = 0$}

Now, let us calculate $J^{0i}$ given by (\ref{angular time momentum}). First notice that the above choice $2 \psi \simeq \rho^{-3}$ means that (\ref{delta-ds2}) becomes
 \begin{equation}\label{delta-ds2-T0}
\delta ds^2|_{T = T_0} = 3 a \sqrt{r_0} \,  \frac{n_i d \rho_i}{\rho^{3/2}}  (n_2 \, d \rho_1 - n_1 \, d \rho_2), 
\end{equation}
which gives for $\delta g_{ij}$
\begin{equation}\label{delta-g}
\delta g_{kj}|_{T = T_0} = \frac{3}{2} \, a \sqrt{r_0} \, \rho^{-7/2} I_{kj}
\end{equation}
where $I_{kj}$ is the matrix
\begin{equation} \label{I-matrix}
I_{kj} = \left(
\begin{array}{ccc}
2 \rho_1 \rho_2 & \quad \rho_2^2 - \rho_1^2 & \quad \rho_2 \rho_3 \\
\rho_2^2 - \rho_1^2 & \quad - 2 \rho_1 \rho_2 & \quad  - \rho_1 \rho_3 \\
\rho_2 \rho_ 3 & \quad - \rho_1 \rho_3 & \quad 0 
\end{array} \right).
\end{equation}
Then from (\ref{delta-g}) and (\ref{I-matrix}) it comes straight away $\delta g|_{T = T_0} \equiv \delta_{ij} \delta g_{ij}=0$. 
Therefore, since $P^{i}=0$, expression (\ref{angular time momentum}) for $J^{0i}$ simplifies to
\begin{equation}\label{angular time momentum-2}
J^{0i} =  \frac{\rho^2}{16\pi} \int (\delta g_{ij} - \rho_i \partial_k \delta g_{kj}) n_j d \Omega, 
\end{equation}
with $d \Omega \equiv \sin \theta d \theta d \Phi$. Then, from (\ref{delta-g}) and (\ref{I-matrix}) $\delta g_{ij} n_j$ depend on $\Phi$ like
\begin{equation}
\delta g_{1j} n_j \propto \sin \Phi, \qquad \delta g_{2j} n_j \propto \cos \Phi,  \qquad \delta g_{3j} n_j =0,  
\end{equation}
from which we obtain right away $\int \delta g_{ij} n_j d \Omega = 0$. Thus (\ref{angular time momentum-2}) reduces to
\begin{equation}\label{angular time momentum-3}
J^{0i} =  -  \frac{\rho^3}{16\pi} \int n_i n_j \partial_k \delta g_{kj}  d \Omega. 
\end{equation}

The contracted divergence $n_j \partial_k \delta g_{kj}$ involves the factor $n_j \partial_k I_{kj}$, whose value is trivially:
\begin{equation} \label{div-I-matrix}
n_j \partial_k I_{kj} = (n_1, n_2, n_3) \cdot (5 \rho_2, - 5 \rho_1, 0) = 0.
\end{equation}

Thus
\begin{equation} \label{div-delta-metrica}
n_j \partial_k \delta g_{kj} = - \frac{21}{4} \, a \sqrt{r_0} \rho^{-9/2} \, n_j n_k I_{kj}.
\end{equation}

Then the calculation gives
\begin{equation} \label{nn-hat-I-matrix}
n_j n_k {I}_{kj} = 0
\end{equation}
and consequently (\ref{angular time momentum-3}) becomes
\begin{equation} \label{J0i-zero}
J^{0i} = 0.
\end{equation}
%


\subsection{Concluding remarks}

Notice that the $J^{ij}$ value obtained (see (\ref{3momenta-zero})) is independent of the angular 3-mo\-men\-tum origin since, because of  
(\ref{delta-ds2-punt}) and $2 \psi = \rho^{-3}[1 + {\cal O}(\rho^{-n})$] with $n>0$, we have straightforwardly,  for $T = T_0$,   $\int \delta \dot{g}_{ij} d \Sigma_j  = 0$, which  is the necessary and sufficient condition for this independence, as it can be easily seen from (\ref{angular momentum}).

Otherwise, this kind of decay of $\psi$ in $\rho$ makes, as announced, that $\delta ds^2$ in (\ref{metric-split-2}) fulfills the asymptotic (b) condition faster than $\rho^{-1}$.

All in all, choosing an arbitrary constant time value $T_0$ and fixing the coordinates in the corresponding slice (that is, choosing $C = 3 \sqrt{r_0} \, T_0/2$, in (\ref{rrhoT3})) we have obtained the 4-momenta $P^{\alpha}$ and  $J^{\alpha\beta}$ for the linearized Kerr metric (\ref{KLT-Lemaitre2}). 
Their values,  given by (\ref{3momenta-zero}), (\ref{E0}) and (\ref{J0i-zero}), vanish, independently  of the chosen constant time $T_0$. In all, it should be emphasized that not just one coordinate system is constructed in which the momenta vanish at any time $T$, but that a whole family of coordinates is needed, one for each choice of a hypersurface $T = T_0$, such that the momenta vanish only for $T = T_0$ in the respective coordinate system.

Notice that coordinates $(T, \rho_i)$, in which we have calculated both vanishing 4-momenta, $P^{\alpha}$ and $J^{\alpha\beta}$, are according to the definition given in Section \ref{sec-2} intrinsic coordinates: they are Gauss coordinates which become asymptotically rectilinear ones (see (\ref{K-simple})-(\ref{delta-ds2})), where, as we have seen, the corresponding angular 3-momentum, $J^{ij}$, vanishes irrespective of the origin of 3-momentum we take. 

In other words, the two intrinsic 4-momenta, $P^ {\alpha}$ and $J^ {\alpha\beta}$, for the linearized Kerr metric, vanish. Thus, in our parlance, this metric is a ``creatable''  one, or otherwise said could in principle arise by a quantum fluctuation from {\em nothing} \cite{Albrow,Tryon,Vilenkin-1985}. In this way, since as explained at the beginning of Section \ref{sec-2} this metric becomes the Lense-Thirring one for large enough values of the radial coordinate, we have also proved that the possible ``creatable'' character of the Schwarzschild metric found in \cite{Schwarzschild-Creatable} is a linear stable one inside the Kerr metric family, in other words, the vanishing of its 4-momenta would be physically sound.

\vspace{1cm}


{\bf Acknowledgements}  
This work was supported by the Spanish ``Ministerio de Ciencia e Innovaci\'on'', project FIS2010-15492 and Basque Government, project IT592-13 (J. M. Aguirregabiria) and by 
the Spanish ``Ministerio de  Econom\'{\i}a y Competitividad'', MICINN-FEDER project FIS2012-33582 (R. Lapiedra and J. A. Morales-Lladosa).


\bibliography{apssamp}

\end{document}